\documentclass{Interspeech}

\interspeechcameraready

\usepackage{cite}
\usepackage{amsmath,amssymb,amsfonts}
\usepackage{algorithmic}
\usepackage{graphicx}
\usepackage{textcomp}
\usepackage[table]{xcolor}
\usepackage{url}
\usepackage{bm}
\usepackage{booktabs}
\usepackage{multirow,multicol}
\usepackage{enumitem}
\usepackage{etoolbox}
\usepackage{hyperref}

\usepackage{bibspacing}  

\newcommand{\unnumberedparagraph}[1]{%
  \par\addvspace{1pt}
  \noindent\textbf{#1}
  \ignorespacesafterend
}
\title{vec2wav 2.0: Advancing Voice Conversion via Discrete Token Vocoders}

\author[affiliation={1,2}]{Yiwei}{Guo}
\author[affiliation={1,2}]{Zhihan}{Li}
\author[affiliation={1,2}]{Junjie}{Li}
\author[affiliation={1,2}]{Chenpeng}{Du}
\author[affiliation={1,2}]{Hankun}{Wang}
\author[affiliation={3}]{Shuai}{Wang}
\author[affiliation={1,2}]{Xie}{Chen}
\author[affiliation={1,2}]{Kai}{Yu}

\affiliation{}{X-LANCE Lab, MoE Key lab of Artificial Intelligence, School of Computer Science, Shanghai Jiao Tong University}{China}
\affiliation{}{Jiangsu Key Lab of Language Computing}{China}
\affiliation{}{Shenzhen Research Institute of Big Data}{China}
\email{yiwei.guo@sjtu.edu.cn, kai.yu@sjtu.edu.cn}
\keywords{Voice conversion, discrete speech token, speech self-supervised model, vocoder, speech re-synthesis}

\usepackage{comment}

\begin{document}

\maketitle

\begin{abstract}
We propose a new speech discrete token vocoder, vec2wav 2.0, which advances voice conversion (VC).
We use discrete tokens from speech self-supervised models as the content features of source speech, and treat VC as a prompted vocoding task.
To amend the loss of speaker timbre in the content tokens, vec2wav 2.0 utilizes the WavLM features to provide strong timbre-dependent information.
A novel adaptive Snake activation function is proposed to better incorporate timbre into the waveform reconstruction process.
In this way, vec2wav 2.0 learns to alter the speaker timbre appropriately given different reference prompts.
Also, no supervised data is required for vec2wav 2.0 to be effectively trained.
Experimental results demonstrate that vec2wav 2.0 outperforms all other baselines to a considerable margin in terms of audio quality and speaker similarity in English and cross-lingual any-to-any VC.
Ablation studies verify the effects made by the proposed techniques.
\end{abstract}

\section{Introduction}
{\let\thefootnote\relax\footnotetext{\vspace{-0.2in}{Kai Yu is the corresponding author.}}}
Discretizing speech into ``tokens'' has prevailed in speech generative tasks, such as text-to-speech (TTS)~\cite{VQTTS,valle,du2024unicats,facodec}, in the era of large language models (LLMs). 
However, the potential of discrete speech tokens in voice conversion (VC) has not been fully mined, which typically aims to convert source speech into target timbre from reference speech.
Speech discrete tokens can be roughly divided into acoustic tokens and semantic tokens~\cite{yang2024towards}.
Although general-purpose acoustic tokens~\cite{encodec,kumar2024high} reconstruct speech signals well, 
they lack the ability of VC because all aspects of information in speech are mixed and retained together. 
Semantic tokens usually come from speech self-supervised (SSL) models~\cite{vq-wav2vec,hsu2021hubert,baevski2020wav2vec,chen2022wavlm} that emphasize on content-related information.
No matter whether timbre is intentionally or unintentionally removed in these tokens, they can act as content representations and thus be utilized in the recognition-synthesis VC paradigm~\cite{huang2022s3prl}.



Among literature, VC methods with a continuous feature space have been researched with depth.
These methods include speech decoupling via autoencoder bottlenecks~\cite{qian2019autovc,qian2020unsupervised,chan2022speechsplit2}, and the adoption of advanced generative algorithms like normalizing flow~\cite{casanova2022yourtts,li2023freevc} and diffusion models~\cite{diffvc, choi2023diff,choi2024dddm}.
After the rise of speech SSL methods, researchers begin to apply SSL features in VC
~\cite{huang2022s3prl,hussain2023ace,knnvc,softvc,choi2023nansy,neekhara2023selfvc,qian2022contentvec}
where the rich phonetic content information from SSL features are utilized.

However, VC with continuous features is hard to cooperate with LLMs, thus an isolated step from other speech-related tasks.
Discrete speech tokens can also serve as content representations, thus VC can be treated as a speech re-synthesis task~\cite{polyak21}.
Recently, discrete SSL features are increasingly explored in VC to retain phonetic content while discarding most acoustic details~\cite{polyak21,uuvc,vectokvc+}.
There also exist researches on decoupling speech tokens that also facilitate VC, such as SSVC~\cite{SSVC} and FACodec~\cite{facodec}.
Nevertheless, the performance of those VC methods is still limited compared to continuous state-of-the-arts.
Also, excessive design of speaker disentanglement in the discrete tokens may cause a negative impact on other paralinguistic information that needs to be preserved, such as prosody.

Instead of pursuing perfect disentanglement in tokens, a different approach is to enhance the timbre controllability in discrete token vocoders.
A typical instance is the idea of ``prompted vocoders'' proposed by CTX-vec2wav~\cite{du2024unicats} which is later verified in VC~\cite{li2024sef}.
In CTX-vec2wav, timbre information is injected using a reference prompt.
By its position-agnostic cross-attention mechanism, timbre in the mel-spectrogram prompts can be effectively incorporated into the process of speech re-synthesis than only using a time-invariant speaker embedding vector~\cite{li2024sef}. 
This indicates the larger potential of performing VC through discrete token vocoders.

In this study, we make key improvements upon this framework that significantly boost the effect of acoustic prompts as the source of timbre information.
Advanced SSL features are utilized for providing discriminative timbre representation.
Most notably, we propose a novel adaptive Snake activation function where the magnitude and frequency of the sinusoidal functions are both controlled by the target speaker's timbre features.
This makes the intrinsic periodical properties in the generated signal highly sensitive to the provided timbre features.
The resulting model, vec2wav 2.0, is then a discrete token vocoder with strong timbre controlling abilities while retaining the content and styles from the content discrete tokens.
In general, vec2wav 2.0 has the following advantages:
\begin{itemize}[leftmargin=3mm]
\setlength{\itemsep}{0pt}
\item \textbf{Unity}. vec2wav 2.0 unifies speech discrete token re-synthesis and VC into the same framework of prompted vocoders.
\item \textbf{Simplicity}. vec2wav 2.0 does not need any labeled data to train. The only data assumption is utterances are segmented into single-speaker ones. 
The training criterion is also simple enough, without additional losses for decoupling.
\item \textbf{Competitiveness}. vec2wav 2.0 achieves superior any-to-any VC performance even compared to continuous VC methods and industry-level VC methods. 
Moreover, vec2wav 2.0 exhibits notable cross-lingual VC performance despite being trained only on English data.
\item \textbf{New Paradigm}. vec2wav 2.0 proves that speaker timbre can be almost manipulated solely by vocoders even if the speech tokens are not perfectly speaker-decoupled.
This may simplify the paradigm of the LLM-based TTS world nowadays.

\end{itemize}
Audio demos and source code are available online\footnote{\scriptsize{https://cantabile-kwok.github.io/vec2wav2/}}.

\section{vec2wav 2.0: Prompted Token Vocoder}

\subsection{System Overview}

\begin{figure}
    \centering
    \includegraphics[width=0.99\linewidth]{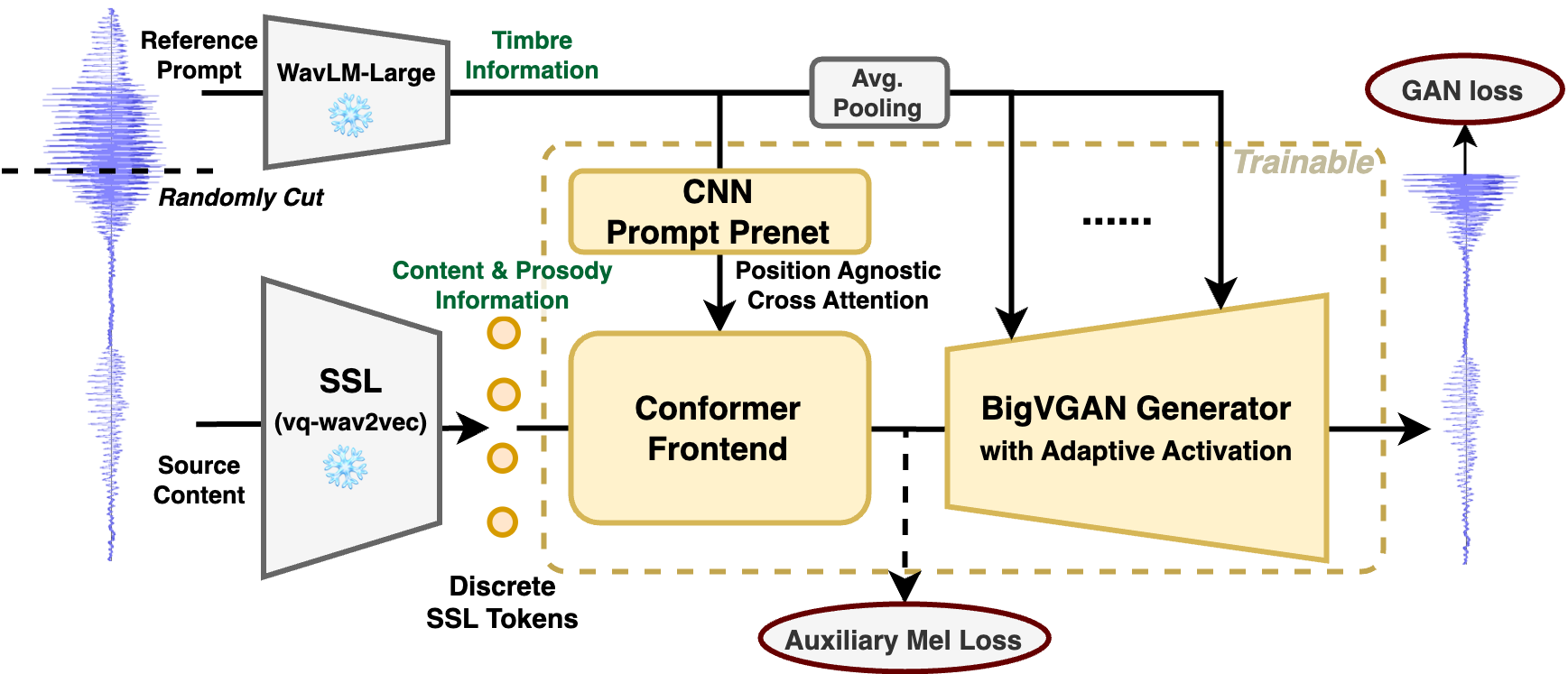}
    \caption{Architecture overview of vec2wav 2.0.}
    \vspace{-0.2in}
    \label{fig:main}
\end{figure}

We design vec2wav 2.0 to be a prompted discrete token vocoder as shown in Fig.\ref{fig:main}.
The overall architecture inherits the frontend-generator framework of CTX-vec2wav~\cite{du2024unicats}, where the input discrete speech tokens are first fed to a Conformer-based frontend module to soften the discreteness, before a vocoder generator that finally outputs the realistic waveforms.
The acoustic prompt brings sufficient timbre information into the process of speech re-synthesis.
We first extract prompt embeddings through a pretrained WavLM model, then use a convolutional neural network (CNN) pre-net to process the hidden embeddings.
In the frontend module, the prompt embeddings are utilized by the position-agnostic cross-attention mechanism~\cite{du2024unicats,li2024sef}, which does not apply positional encoding to the query sequence.
This special cross attention mechanism simulates shuffling the query sequence and inherently breaks the local patterns in the reference prompt, e.g. linguistic and prosodic features, which enables more accurate learning of target timbre as some global information.

After timbre is preliminarily merged into the frontend, we design an adaptive BigVGAN~\cite{lee2023bigvgan} generator to further incorporate the timbre embedding in waveform generation. 
The core component of this adaptive generator is a novel adaptive Snake activation function, which will be illustrated in Section \ref{sec:snake}.

\subsection{Adaptive Snake Activation}
\label{sec:snake}
The Snake activation function is proposed in \cite{snake} for modeling periodical inductive bias, which is then adopted in the BigVGAN vocoder to achieve state-of-the-art performance.
This activation function can be represented as $f_\theta(x)=x+\frac1\beta\sin^2(\alpha x)$. The learnable parameters $\theta=\{\alpha,\beta\}$ are designed to control the frequency and magnitude respectively, and $f_\theta$ can operate on each input channel independently, i.e. different $\theta$ for each input channel.

As this Snake activation can subtly capture the periodical pattern in the speech signals, we propose to inject more information from the target speaker timbre. Let $\bm s\in\mathbb R^d$ be some representative speaker embedding extracted from the target speaker, we design an adaptive Snake activation where the frequency and magnitude of sinusoidal function are both affected by $\bm s$:
\begin{align}
    T(\bm s)&=\tanh(W\bm s+\bm b)\\
    f_{\theta}(\bm x, \bm s)&=\bm x+\frac1{\bm\beta+\frac12 T(\bm s)}\sin^2\left[(\bm\alpha + T(\bm s))\bm x\right]\label{eq:adaptive-snake}
\end{align}
where $T$ is a linear transform followed by $\tanh$ activation, and operations in \eqref{eq:adaptive-snake} are all element-wise. $T(\bm s)$ is discounted by $1/2$ on the magnitude part for numerical stability. To save parameters, we apply the same $T$ transformation to both magnitude and frequency. 
In this way, the learnable parameter for each adaptive Snake is $\theta=\{\bm\alpha,\bm\beta,W,\bm b\}$, and the target timbre information can be effectively injected in every layer of the vocoder via adaptive activations, which strengthens the timbre controllability to a considerable extent.

\begin{figure}
    \centering
    \includegraphics[width=0.9\linewidth]{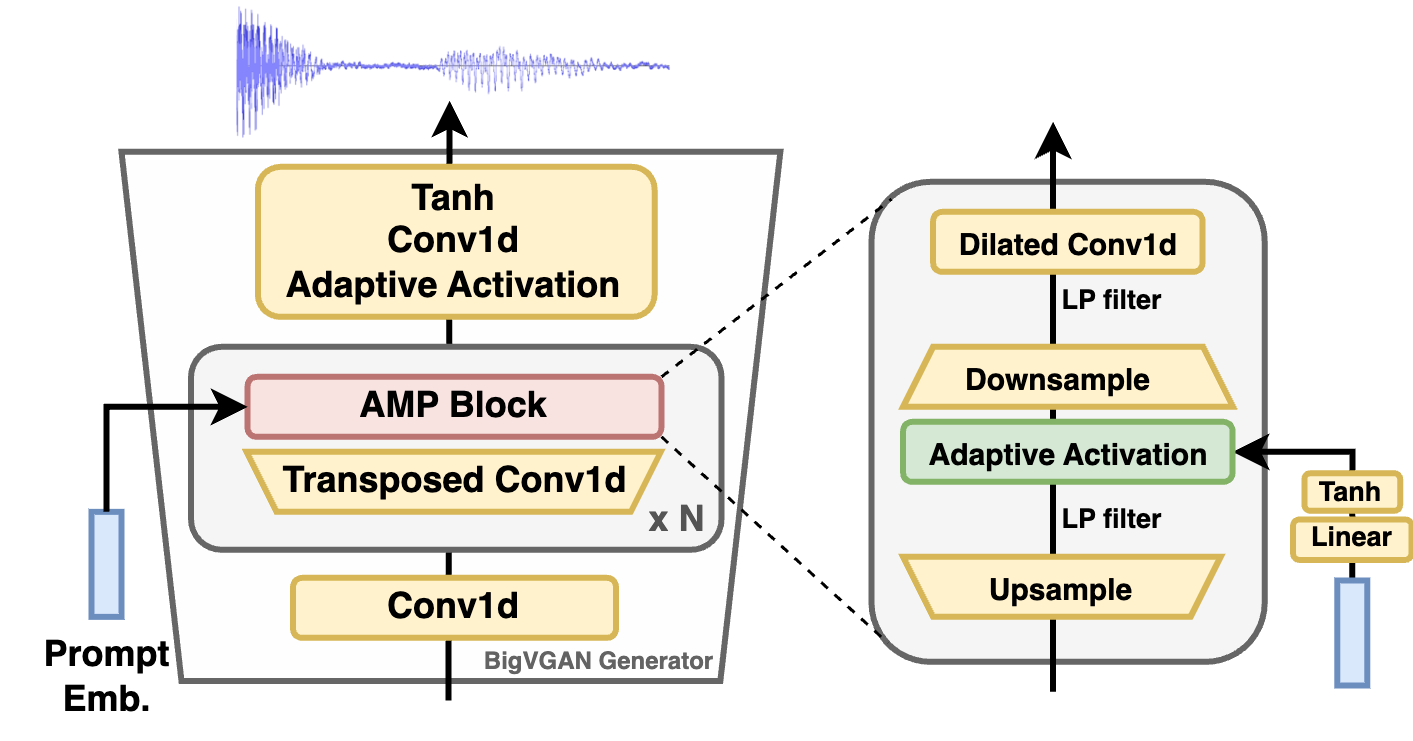}
    \vspace{-0.1in}
    \caption{Detailed architecture of BigVGAN generator with proposed adaptive Snake activations.}
    \vspace{-0.1in}
    \label{fig:bigvgan-detail}
\end{figure}

Here in vec2wav 2.0, the prompt embeddings are first mean-pooled to form a single vector that averages out linguistic details and preserves global timbre, then inserted to every adaptive activation layer in the BigVGAN generator.
Fig.\ref{fig:bigvgan-detail} illustrates the detailed architecture of adaptive BigVGAN generator.
The input hidden states are iteratively upsampled by transposed convolutions and transformed by anti-aliased multi-periodicity composition (AMP) blocks. 
Each AMP block receives an additional prompt embedding that is fed to the adaptive Snake activation layer for timbre control.
Low-pass (LP) filters are applied after each upsampling and downsampling operation to prevent aliasing~\cite{lee2023bigvgan}.
The hidden states are recovered to sampling points after a final adaptive Snake and convolution block. 

\subsection{Content and Prompt Features}
Both the content and prompt inputs to vec2wav 2.0 are SSL features with different goals:
the input tokens should have as less timbre as possible, while the prompt features should contain sufficient and clear timbre information to aid reconstruction.

\unnumberedparagraph{Content Features}~~
We use the off-the-shelf vq-wav2vec~\cite{vq-wav2vec} SSL model for extracting the discrete content representation to be re-synthesized.
The discrete tokens are extracted from the quantizer output before the feature aggregator, which is a two-group integer index sequence. 
We favor this representation because a lot of speaker timbre information is removed due to the contrastive criterion, while most of the phonetic pronunciation and prosody are retained~\cite{DSETTS}.
Also, compared to HuBERT~\cite{hsu2021hubert}-style Transformer SSL models, vq-wav2vec is free of manual clustering and is also fully convolutional with a certain receptive field.
This produces a representation that is unaware of the total sequence length, keeping consistent results for a given window.
This consistency also shows potential for cross-lingual conversion, as its language-agnostic property has been successfully applied in multilingual TTS~\cite{limmits23}.
Although there exists measurable speaker timbre leakage in the discrete tokens~\cite{superb,TNVQTTS,DSETTS}, the vec2wav 2.0 vocoder exhibits strong timbre controllability, so that competitive VC can still be achieved.

\unnumberedparagraph{Prompt Features}
~~Following CTX-vec2wav, the reference prompt segment is randomly cut from the current utterance, to maintain the same speaker identity without labeled data. Instead of using mel-spectrogram to provide timbre information from the reference prompt, we use a pretrained WavLM~\cite{chen2022wavlm} model as a timbre feature extractor owing to its widely-verified advantage on speaker verification~\cite{jung2024espnet,superb}. We freeze the WavLM model in training and only use the output feature at a certain location of its Transformer blocks.
In practice, we use the 6th layer of WavLM-Large model as early layers are proven to contain rich timbre information~\cite{knnvc}.


\subsection{Discriminators and Training Criterion}
We inherit the multi-scale discriminators (MSD) and multi-period discriminators (MPD) from HifiGAN~\cite{kong2020hifigan}. These discriminators are jointly trained with the generator to distinguish fake signals from real ones in multiple scales and periods. 
With the generator adversarially trained to fool the discriminators, we achieve high-fidelity speech re-synthesis and VC results.
Different from some current VC models that often suffer from audio quality issues, vec2wav 2.0 ensures the audio quality of speech signals by GAN training.

The training criteria include the auxiliary mel prediction loss 
and all the other GAN losses 
from HifiGAN.
The auxiliary mel prediction loss is an L1 loss between the ground truth mel-spectrograms and predicted ones that come from linear projections after the Conformer frontend, to warm up the whole model.
This loss is weighted with a certain coefficient, and we cancel it after warming up, following \cite{VQTTS,du2024unicats}.

\subsection{Any-to-Any Voice Conversion}
Although not directly optimized for VC, vec2wav 2.0 still has strong conversion ability due to its effectiveness on incorporating target speaker timbre.
The content features retain most of the phonetic and prosodic information while losing much speaker identity, while the speaker timbre is controlled by the reference prompt.
Therefore, we can achieve VC simply by using the target speaker's reference speech as the prompt input.
This method naturally supports any-to-any VC because the content and prompt features are both acquired by SSL models trained on data with enough speaker variations.

Moreover, as both the cross attention mechanism and the adaptive Snake activation are position agnostic, the ordering of the prompt features plays minimal role in timbre control.
This allows cross-lingual VC where target speakers may come from unseen languages, since almost all linguistic-relevant patterns are broken by these position-agnostic operations.
As long as the global traits are apparent enough in the WavLM features, speaker timbre can be successfully transferred, even if the model is not trained on multilingual data.

\section{Experiments}
\subsection{Data and Model Setup}
We use all the train splits of LibriTTS~\cite{libritts}, an English corpus with 585 hours of 24kHz speech data spoken by around 2500 speakers, to train vec2wav 2.0.
We only keep utterances from 6s to 30s to ensure proper prompt lengths.
The resulting training set has around 360 hours.
The prompt segment is cut starting from a random point within 1 second of either the beginning or the end of an utterance, extending inward towards the middle, with its length randomly sampled between one third and one half of the original utterance's duration.
In this way, a reasonable range of prompt lengths is covered in training, and vec2wav 2.0 learns to handle short reference lengths well.

We use the k-means version of official vq-wav2vec model\footnote{\scriptsize{https://github.com/facebookresearch/fairseq/tree/main/examples/wav2vec}} to extract content tokens from source speech.
As this model adopts grouped vector quantization, we concatenate the code-vectors corresponding to each group before feeding the Conformer frontend.
The input to the frontend is thus a 512-dimensional sequence in 10ms strides.
The prompt embeddings are extracted from official WavLM-Large\footnote{\scriptsize{https://github.com/microsoft/unilm/tree/master/wavlm}} at the 6th layer.

The Conformer frontend of vec2wav 2.0 contains 2 Conformer blocks, where each of the self and cross attention modules has 2 heads and 184 attention dimensions.
The prompt prenet has four CNN blocks with scaled residual connections, where the hidden dimensions are 128, 256 and 512 before being fed to cross attentions.
The resulting generator model has 40.3M parameters.

The whole model is trained for 1 million steps on 4 NVIDIA A10 GPUs with a max batch size of 36s speech data per device. Other hyper-parameters follow CTX-vec2wav~\cite{du2024unicats}.

\subsection{English Any-to-Any VC}
\label{sec:en}

We conduct English any-to-any VC comparisons using the unseen speakers in the LibriTTS test-clean split.
We randomly select 10 speakers, from each of whom 2 utterances are chosen to be the source utterances. Another 10 speakers are selected as target speakers with one 3-second reference utterance for each.
This yields a test set of 200 any-to-any VC cases.

To comprehensively evaluate the performance of VC systems, we employ a range of objective and subjective metrics:
\begin{enumerate}[leftmargin=3mm]
    \item \textbf{Quality and intelligibility}: We use the subjective naturalness MOS (NMOS) and word error rate (WER) between ground truth and recognized texts.
    The NMOS tests require listeners to rate the utterances by quality and naturalness ranging from 1 to 5.
    WERs are computed using NeMo ASR\footnote{\scriptsize{https://huggingface.co/nvidia/stt\_en\_fastConformer\_transducer\_large}}.
    \item \textbf{Speaker similarity}: We conduct similarity MOS (SMOS) tests and compute speaker embedding cosine similarity (SECS).
    Listeners in SMOS tests are asked to rate timbre similarity between reference and synthesized items in 1-5 scale.
    SECS is computed via Resemblyzer\footnote{\scriptsize{https://github.com/resemble-ai/Resemblyzer}} where speaker embeddings are extracted by a verification model for computing cosine similarity in range of -1 to 1.
    \item \textbf{Prosody preservation}: We additionally measure the correlation coefficient of pitch contours (P.Corr) between the source speech and converted speech.
    This is also an important metric in VC because ideal VC systems should preserve prosodic variations in source speech while transferring timbre attributes.
    The value range is -1 to 1, with higher values indicating better preservation.
\end{enumerate}

\begin{table}[]
\centering
\caption{Evaluation of English any-to-any VC}
\vspace{-0.1in}
\label{tab:en}
\resizebox{0.9\columnwidth}{!}{
\begin{tabular}{@{}lccccc@{}}
\toprule
\textbf{Model} & \textbf{NMOS$\uparrow$} & \textbf{WER$\downarrow$} & \textbf{SMOS$\uparrow$} & \textbf{SECS$\uparrow$} & \textbf{P.Corr$\uparrow$} \\ \midrule
Source GT & 4.70$\pm$0.09 & 1.10 & - & - & 1.000 \\ \midrule
\multicolumn{6}{l}{\textit{Academic baselines}} \\
YourTTS~\cite{casanova2022yourtts} & 3.77$\pm$0.12 & 3.95 & 3.47$\pm$0.10 & 0.766 & \textbf{0.758} \\
DiffVC~\cite{diffvc} & 4.08$\pm$0.12 & 6.33 & 3.75$\pm$0.10 & 0.855 & 0.153 \\
Diff-HierVC~\cite{choi2023diff} & 4.23$\pm$0.10 & 1.59 & 4.10$\pm$0.09 & 0.828 & {0.740} \\
UUVC~\cite{uuvc} & 3.72$\pm$0.14 & 2.19 & 3.27$\pm$0.12 & 0.753 & 0.336 \\
FACodec~\cite{facodec} & 4.02$\pm$0.13 & \textbf{1.15} & 3.77$\pm$0.10 & 0.817 & 0.517 \\
\rowcolor{gray!20} 
vec2wav 2.0 & \textbf{4.51$\pm$0.09} & 3.29 & \textbf{4.46$\pm$0.08} & \textbf{0.886} & 0.722 \\ \bottomrule
\end{tabular}
}
\resizebox{0.9\columnwidth}{!}{
\begin{tabular}{@{}lccccc@{}}
\toprule
\textbf{Model} & \textbf{NMOS$\uparrow$} & \textbf{WER$\downarrow$} & \textbf{SMOS$\uparrow$} & \textbf{SECS$\uparrow$} & \textbf{P.Corr$\uparrow$} \\ \midrule
Source GT & 4.53$\pm$0.10 & 1.10 & - & - & 1.000 \\ \midrule
\multicolumn{6}{l}{\textit{Industry-level baseline}} \\
CosyVoice~\cite{du2024cosyvoice} & 4.20$\pm$0.11 & 1.48 & 4.23$\pm$0.10 & 0.871 & 0.617 \\
\rowcolor{gray!20} 
vec2wav 2.0~~~~~~~~~& 4.22$\pm$0.11 & 3.29 & 4.29$\pm$0.11 & 0.886 & 0.722 \\ \bottomrule
\end{tabular}
}
\vspace{-0.1in}

\end{table}

We compare vec2wav 2.0 with some famous VC models. 
YourTTS~\cite{casanova2022yourtts} is a famous flow-based end-to-end VC model.
DiffVC~\cite{diffvc} and Diff-HierVC~\cite{choi2023diff} promote convertibility via diffusion models.
UUVC~\cite{uuvc} also performs VC by discrete token reconstruction, but incorporates HuBERT tokens and additional prosody predictions.
FACodec~\cite{facodec} is a speech codec based on supervised decoupling of content, prosody, timbre and detail information.
FACodec is capable of converting voices by simply replacing the speaker embedding into the target speaker and then decoding into waveform.
We discard the \textit{detail} tokens in FACodec for VC since we find these tokens still contain considerable speaker information that harms VC performance.
We also conduct a separate listening test with a strong industry-level baseline, CosyVoice~\cite{du2024cosyvoice}, which is trained on massive data.
Its VC ability is based on the resynthesis of its supervised speech tokenizer using flow matching.
We use the official checkpoints for all baselines.
Note that the training data in all baselines either includes LibriTTS or is magnitudes larger (e.g. FACodec, CosyVoice), so the comparisons are fair enough.

Table \ref{tab:en} presents the comparison results.
``Source GT'' means source utterance recordings, and MOS values are reported with 95\% confidence intervals.
It is clear that vec2wav 2.0 achieves significantly higher synthesis quality and speaker similarity than all the academic baselines.
Compared to CosyVoice, vec2wav 2.0 still owns slightly better performance in naturalness, similarity and prosody preservation, although the training data size is 300 times smaller.
Its pitch correlation is also at a high level\footnote{ Note that pitch correlation is less meaningful if speaker similarity is low.}.
While the WER of vec2wav 2.0 is not the lowest, it remains acceptable. This is mostly due to the quantization errors inherent in the vq-wav2vec model itself.

\vspace{-0.05in}
\subsection{Cross-Lingual Any-to-Any VC}
\begin{table}[]
\centering
\caption{Evaluation of cross-lingual any-to-any VC}
\vspace{-0.1in}
\label{tab:cross}
\resizebox{0.9\columnwidth}{!}{
\begin{tabular}{@{}lccccc@{}}
\toprule
\textbf{Model} & \textbf{NMOS$\uparrow$} & \textbf{WER$\downarrow$} & \textbf{SMOS$\uparrow$} & \textbf{SECS$\uparrow$} & \textbf{P.Corr$\uparrow$} \\ \midrule
Source GT & 4.73$\pm$0.07 & 1.10 & - & - & 1.000\\ \midrule
YourTTS & 3.57$\pm$0.09 & 4.90 & 3.38$\pm$0.11 & 0.772 & \textbf{0.731} \\
Diff-HierVC & 4.08$\pm$0.08 & \textbf{1.59} & 4.14$\pm$0.09 & 0.805 & 0.728 \\
\rowcolor{gray!20} 
vec2wav 2.0 & \textbf{4.47$\pm$0.07} & 3.39 & \textbf{4.33$\pm$0.07} & \textbf{0.846} & 0.684 \\ \bottomrule
\end{tabular}
}
\vspace{-0.2in}
\end{table}

To verify the cross-lingual VC ability of vec2wav 2.0, 
we use the same set of English source utterances in Table \ref{tab:en}, but convert to target speakers in other languages.
We collect reference utterances from five languages\footnote{ Spanish, German, Dutch, Italian, French.} in MLS~\cite{MLS}.
The test set is the full combination of source and target utterances.
For each of those languages, one male and one female speaker are randomly chosen as target speakers, and one reference utterance for each target speaker is sampled.
We compare vec2wav 2.0 with the famous cross-lingual VC model YourTTS that is trained on multilingual data, and also Diff-HierVC which is a competitive academic baseline in Table \ref{tab:en}.
We conduct subjective and objective evaluations in the same way as Section \ref{sec:en}. 

Table \ref{tab:cross} shows the results.
Although not trained on multilingual data, vec2wav 2.0 consistently outperforms YourTTS and Diff-HierVC in speaker similarity and quality with a significant margin.
The WER and P.Corr comparisons show a similar conclusion with Table \ref{tab:en} that vec2wav 2.0 possesses a decent level of intelligibility and prosody preservation, although not the best.
Therefore, it is demonstrated that vec2wav 2.0 performs competitive conversions, regardless of the languages of references.

\vspace{-0.1in}
\subsection{Ablation Study}
\begin{figure}
    \centering
    \includegraphics[width=0.8\linewidth]{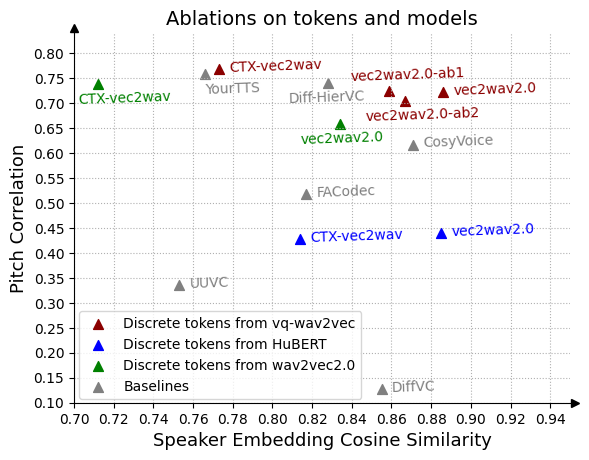}
    \vspace{-0.1in}
    \caption{Objective SECS and P.Corr comparisons with varied input tokens and models. Perfect VC systems should lie on the top right corner.}
    \vspace{-0.2in}
    \label{fig:ablation}
\end{figure}
We also conduct ablation studies on different input SSL discrete tokens and vocoder architectures.
Apart from vq-wav2vec, we train CTX-vec2wav (our predecessor) and vec2wav 2.0 on HuBERT tokens and wav2vec 2.0~\cite{baevski2020wav2vec} tokens.
The HuBERT tokens are obtained by 2048-centroid clustering on the output of the last layer.
The wav2vec 2.0 tokens are considered the quantizer output before the Transformer, with 2 codebook groups each with 320 codes.

To compare architectures, we additionally train two variants of vec2wav 2.0 on vq-wav2vec inputs: \textbf{vec2wav 2.0-ab1} that replaces the adaptive Snake activations in BigVGAN by the original Snakes; and \textbf{vec2wav 2.0-ab2} that further replaces BigVGAN with HifiGAN.
Thus the comparison between vec2wav2.0 and ``ab1'' indicates the effect of adaptive Snake activation, while that between CTX-vec2wav and ``ab2'' shows the difference made by prompt feature and modules.
We present the ablation studies in terms of SECS and P.Corr in Fig.\ref{fig:ablation}, together with the baselines in Section \ref{sec:en}.
It can be found that vec2wav 2.0 obtains consistently large improvements in speaker similarity compared to the predecessor CTX-vec2wav in all the three input SSL tokens, while maintaining comparable pitch preservation.
From the ablation of model architectures, it is obvious that the prompt-related improvements of vec2wav 2.0 make a substantial contribution to speaker similarity, while the adaptive Snake activations further advance the VC performance.
The proposed vec2wav 2.0 with vq-wav2vec tokens is finally nearest to the top right corner of Fig.\ref{fig:ablation}, pushing the frontier of modern VC methods towards ideal voice converters.



\vspace{-0.05in}
\section{Conclusion}

We present a novel VC method, vec2wav 2.0, based on the re-synthesis of speech discrete tokens.
It takes advantage of SSL features in both content and timbre representations and enhances CTX-vec2wav in architectural designs.
The adaptive Snake activation technique is proposed to better incorporate timbre into waveform reconstruction.
The resulting model achieves remarkable performance on intra and cross-lingual VC tasks.
We believe vec2wav 2.0 has promising impacts on the future LLM-based speech generation paradigm.
Future efforts are needed in improving the intelligibility and prosody preservation of the proposed method, and the scaling ability on large-scale in-the-wild datasets needs to be explored.



\bibliographystyle{IEEEtran}
\bibliography{refs}

\end{document}